%% file: JVernet_astroph.tex
\documentclass[a4paper, traditabstract, longauth]{aa} 
\usepackage{txfonts}
\usepackage{graphicx}
\usepackage{placeins}
\usepackage{float}
\usepackage{subfigure}
\usepackage{amsmath}
\usepackage{mathrsfs}
\usepackage{amssymb}
\usepackage{color,epstopdf}
\usepackage{pdflscape}

\newcommand{\src}{MRC\,0316-257}
\newcommand{\surfb}{erg\,s$^{-1}$\,cm$^{-2}$\,arcsec$^{-2}$}
\newcommand{\Msun}{M$_{\odot}$}

\newcommand{\Lya}{Ly$\alpha$}

\newcommand{\CIV}{C{\small IV}}
\newcommand{\HeII}{He{\small II}}

\newcommand{\OII}{[O{\small II}]}
\newcommand{\CIII}{C{\small III}]}

\newcommand{\CO}{CO(8--7)}
\newcommand{\kms}{km\,s$^{-1}$}

\begin{document}

\title{Are we seeing accretion flows in a 250kpc-sized Ly$\alpha$
halo at z=3?\thanks{Based on observations made with ESO Telescopes at
the La Silla Paranal Observatory under program ID 094.B-0699(A)}}

\author{J.~Vernet\inst{1}
\and
M.~D.~Lehnert\inst{2}
\and
C.~De~Breuck\inst{1}
\and 
M.~Villar-Mart\'{i}n\inst{3}
\and
D.~Wylezalek \inst{4}
\and
T.~Falkendal\inst{1,2}
\and 
G.~Drouart \inst{5}
\and
S.~Kolwa\inst{1}
\and
A.~Humphrey \inst{6}
\and
B.~P.~Venemans\inst{7}
\and
F. Boulanger\inst{8,9}}

\institute{European Southern Observatory, Karl-Schwarzschild-Str. 2, D-85748 Garching \email{jvernet@eso.org}
\and
Sorbonne Universit\'es, UPMC, Paris VI,
CNRS UMR 7095, Institut d'Astrophysique de Paris, 98bis bvd Arago,
75014, Paris, France
\and
Centro de Astrobiolog\`{i}a (INTA-CSIC), Carretera de Ajalvir, km 4,
28850 Torrej\`{o}n de Ardoz, Madrid, Spain
\and
Department of Physics and Astronomy, Johns Hopkins University, 3400 N. Charles St, Baltimore, MD, 21218, USA
\and
International Centre for Radio Astronomy Research, Curtin University, Perth WA 6845, Australia
\and
Instituto de Astrofísica e Ci\^{e}ncias do Espa\c{c}o, Universidade do Porto, CAUP, Rua das Estrelas, P-4150-762 Porto, Portugal
\and
Max-Planck Institute for Astronomy, K\"{o}nigstuhl 17, 69117 Heidelberg, Germany
\and
Institut d'Astrophysique Spatiale, 91405 Orsay, Universit\'e Paris
Sud et CNRS, France
\and
Research associate at Institut d'Astrophysique de Paris}

\abstract{Using MUSE on the ESO-VLT, we obtained a 4 hour exposure of the
z=3.12 radio galaxy \src. We detect features down to $\sim$10$^{-19}$
\surfb\ with the highest surface brightness regions reaching more than
a factor of 100 higher. We find \Lya\ emission out to $\sim$250\,kpc
in projection from the active galactic nucleus (AGN). The emission
shows arc-like morphologies arising at 150-250\,kpc from the nucleus
in projection with the connected filamentary structures reaching down
into the circum-nuclear region. The most distant arc is offset by
$\sim$700\,\kms\ relative to circum-nuclear \HeII$\lambda$1640 emission,
which we assume to be at the systemic velocity. As we probe emission
closer to the nucleus, the filamentary emission narrows in projection on
the sky, the relative velocity decreases to $\sim$250\,\kms , and line
full-width at half maximum range from $\sim$300-700\,\kms. From UV line
ratios, the emission on scales of 10s of kpc from the nucleus along
a wide angle in the direction of the radio jets is clearly excited by
the radio jets and ionizing radiation of the AGN. Assuming ionization
equilibrium, the more extended emission outside of the axis of the
jet direction would require 100\% or more illumination to explain the
observed surface brightness. High speed ($\gtrsim$300\,\kms) shocks into
rare gas would provide sufficiently high surface brightness. We discuss
the possibility that the arcs of \Lya\ emission represent accretion
shocks and the filamentary emission represent gas flows into the halo,
and compare our results with gas accretion simulations.}

\keywords{Galaxies: evolution -- Galaxies: high redshift -- Galaxies:
active -- Galaxies: ISM -- Galaxies: halos}

\titlerunning{Ly$\alpha$ halo of MRC\,0316-257}
\authorrunning{Vernet, Lehnert, De Breuck et al.}

\maketitle

\section{Introduction}

High-redshift radio galaxies, quasars \citep{heckman91b, heckman91a},
QSOs \citep{christensen06}, and ``\Lya\ blobs'' \citep{steidel00} exhibit
large, extended \Lya\ emission. High-redshift radio galaxies play
a central role in our evolving understanding of the association between
massive galaxies, their clustered environments, and halo gas. In fact,
the first galaxies at high redshift, z$>$2, where halo gas was detected
in \Lya\ emission over scales of 10-100s kpc, were radio galaxies
\citep{chambers90}. These \Lya\ halos are generally aligned with the radio
jets but extend well beyond the radio lobes \citep{villar-martin03b}. It
is not clear what the association is between the active galactic nuclei
(AGN) and the extended \Lya\ emission. In the circum-galactic environment
within the radius subtended by the radio lobes, the \Lya\ kinematics
are complex, while outside the radio emission, the gas is relatively
quiescent \citep{villar-martin03b}.

The energy sources of \Lya\ nebulae are not well-constrained
\citep{cantalupo16}. All processes that excite the \Lya\ emission
depend on the distribution of the emission line gas, the type of objects
within the nebula, and how far the emission is from sources of ionizing
photons. While ionizing photons from galaxies embedded within the \Lya\
emission is plausible \citep{overzier13}, other sources powering the
\Lya\ emission include ionization by the meta-galactic flux, mechanical
heating, dissipation of potential energy as the gas falls into the
halo, and resonance scattering of \Lya\ and UV continuum photons.
The mechanisms powering the emission are inextricably linked with
the origin of the gas. If the emission is related to outflows, the
excitation of the gas might be due to shocks and ionization by stars or
AGN \citep{swinbank15}. If the gas extends $>$100 kpc, it may be accreting
from the cosmic web or gas instabilities in the halo \citep{maller04}. In
that case, we expect the emission to be due to mechanical heating
associated with accretion shocks \citep{birnboim03} or from dissipation
of potential energy the flows gain as they fall into the potential.

While these general arguments motivated us to obtain deep integral-field
spectroscopy of radio galaxies using MUSE on the ESO-VLT, in the specific
case of \src, we are motivated by its striking \Lya\ morphology.  \src\
is well-studied, massive $\approx$2$\times$10$^{11}$\,\Msun, radio galaxy
\citep{debreuck10}, which lies in a galaxy over-density \citep{kuiper12}.
The \Lya\ emission around \src\ is filamentary on scales of $\sim$250\,kpc
and has a morphology and surface brightness distribution similar to
those seen in galaxy simulations. Simulations of gas accretion imply
that streams will be visible via their \Lya\ emission if we detect down
to a surface brightness of $\approx$10$^{-19}$\,\surfb\ at z$\sim$3
\citep{rosdahl12}. It is not clear if these simulations are capturing all
the physics necessary to model the emission and evolution of the accreting
gas \citep[e.g.,][]{cornuault16}. Observational constraints are needed.
Of course, one must worry about expectation bias \citep{jeng06}. Do we
call what we observe a ``stream'' because its morphology agrees with the
results of simulations? While we are concerned about this bias, we think
it is still instructive to compare our results with those of simulations.

\section{Observations, analysis, and results}\label{sec:obs}

Multi Unit Spectroscopic Explorer \citep[MUSE;][]{bacon10} observations
of \src\ were obtained in service mode between UT January 2015 January
14 and 17.  We used the Wide Field Mode (1$\arcmin$$\times$1$\arcmin$
field of view) with the 2$^{nd}$-order blocking filter resulting in
a wavelength coverage of 480-935 nm. Our eight 900s exposures were
taken at position angles of 0, 90, 180 and 270$^{\circ}$ with a small
pointing offset to mitigate against systematic artefacts.  We processed
the data using MUSE pipeline version 1 \citep{weilbacher12} to produce a
fully-calibrated (wavelength, flux, and astrometry) sky-subtracted data
cube. The measured image quality of the reconstructed white light image is
$\sim$1\arcsec. To preserve any possible extended low surface-brightness
features, we did not use sky-residual cleaning algorithms.

\begin{figure}
\begin{center}
\includegraphics[width=7.5cm]{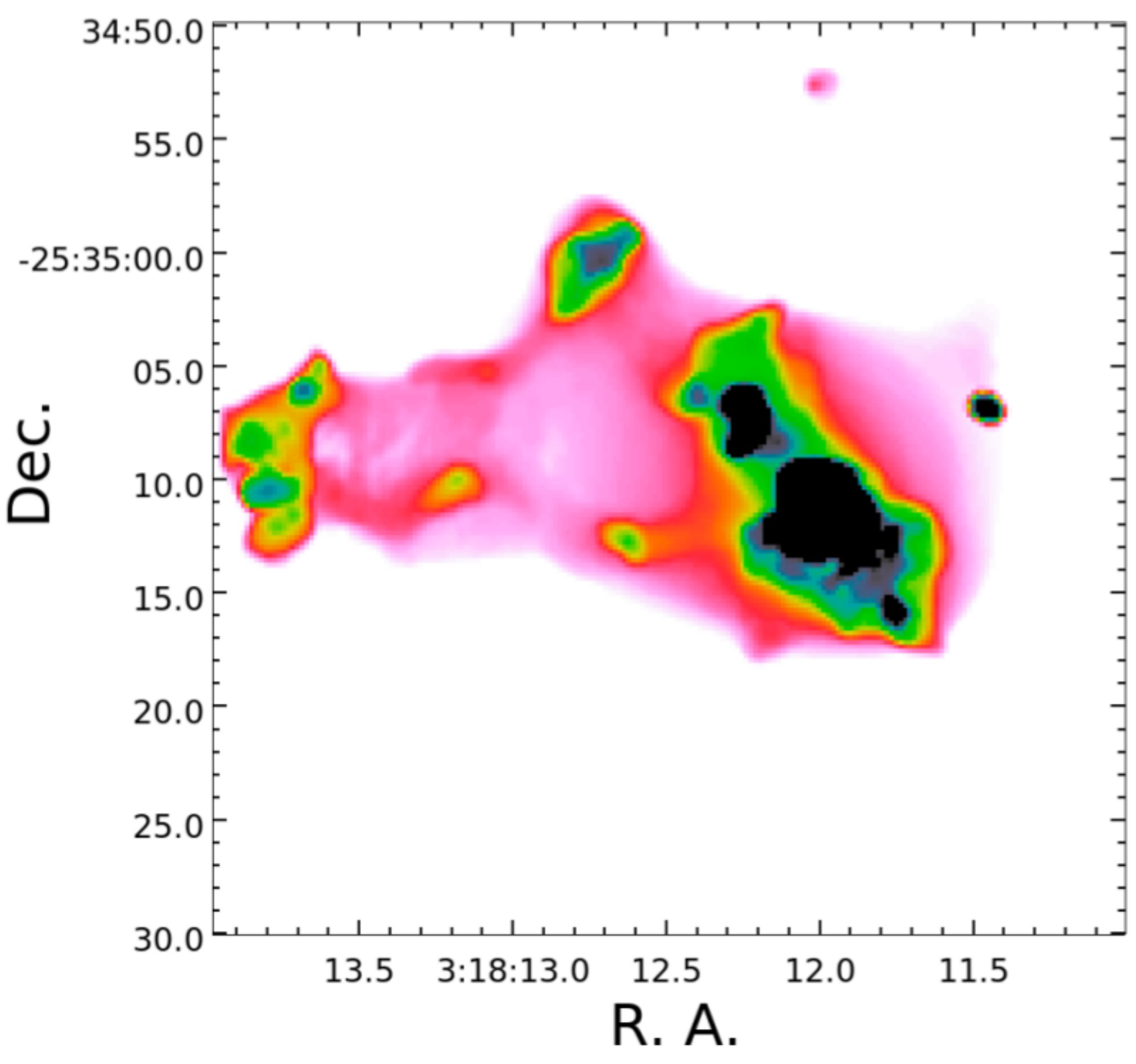}
\end{center}
\caption{A 25$\AA$ wide narrow-band image centered on \Lya\ extracted
from an adaptively smoothed datacube with the algorithm described in
\citet{martin14}.  We used a maximum spatial smoothing scale of 14\arcsec\
and a signal-to-noise ratio threshold of 2.5 in making this image.  This
plot is intended to be illustrative and was not used in the analysis. }
\label{fig:adaptivesmoothing}
\end{figure}

\begin{figure*}
\begin{center}
\includegraphics[width=16.5cm]{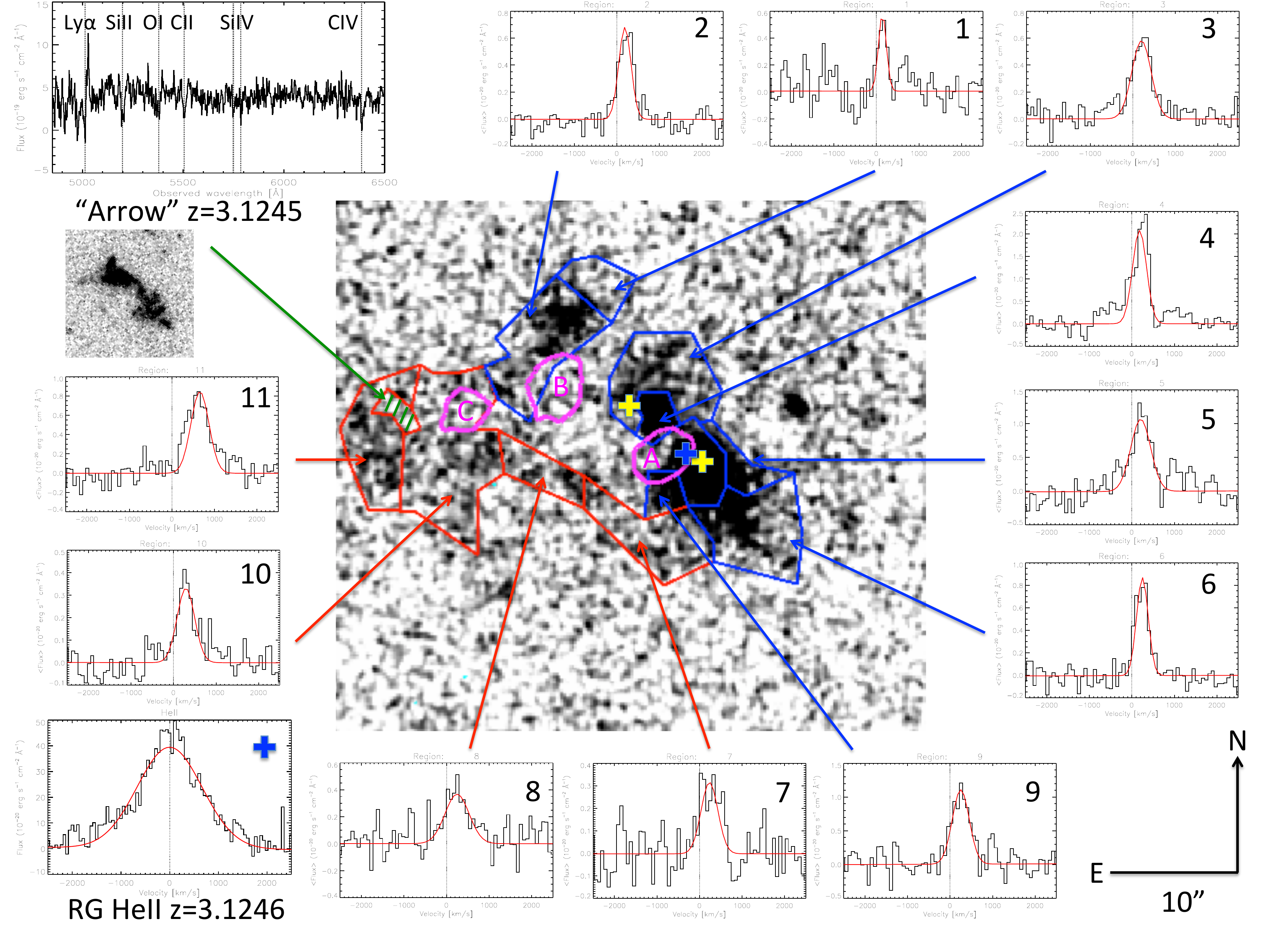}
\end{center}
\caption{Grayscale image of the \Lya\ emission in \src\ (scale and
orientation at the bottom right).  The red and blue lines demarcate
 11 regions with their mean \Lya\ spectra shown in the surrounding
 plots. The regions bounded by blue lines are likely directly
excited by the AGN while the regions bounded by red lines indicate
regions perhaps associated with a ``stream''.  In each of the spectral
plots, we approximate the \Lya\ profile with the best fit Gaussian
(red).  The velocity is relative to systemic (in \kms; indicated by
black vertical line) measured from the \HeII\ profile (bottom, left).
The purple contours show the region of \OII\ emission for 3 foreground
galaxies, labeled A, B, and C, whose redshifts are 0.874, 0.851, and
0.667, respectively.   We indicate the positions of the radio lobes
(yellow crosses) and AGN (blue cross). Top two left panel shows the
	spectrum and HST F814W image of the ``Arrow'' (green hatched region in
the grayscale image).}

\label{fig:imagespectralmontage} 
\end{figure*}

To remove remaining artefacts, we subtracted all continuum sources
from the datacube using a linear interpolation between two 50\AA\
wide bins on both sides of the emission line. The very
extended \Lya\ emission is only marginally detected. To make the
\Lya\ morphology more evident, we implemented our own version of the
algorithm of \citet[][Fig.~\ref{fig:adaptivesmoothing}]{martin14}. To
provide the characteristics of the \Lya\ emission, we extracted and
fit the \Lya\ line from several regions (Table~\ref{tab:specnumbers},
Fig.~\ref{fig:imagespectralmontage}).  The line ratios
of \Lya , \CIV$\lambda\lambda$1548,\,1551, \HeII$\lambda$1640, and
\CIII$\lambda\lambda$1907,\,1909 of the highest surface brightness regions
(3-5) imply they are ionized by the AGN.  Beyond the circum-nuclear
emission, $\ga$10\arcsec\ from the AGN, the \Lya\ emitting regions are
too faint to determine line ratios necessary to constrain the ionization
source.

\input{Results_fit.tex}

We find \Lya\ emission, down to $\approx$10$^{-19}$\,\surfb ,
up to $\sim$35$\arcsec$ from the radio galaxy ($\sim$250\,kpc in
projection). The morphology of the emission, outside of that likely
excited by the AGN, is arc- and stream-like with the higher surface
brightness arcs appearing at $\sim$150-250 kpc (regions 1, 2, and 11)
from the nucleus with the connected filamentary structures reaching
into the circum-nuclear regions (regions 6-8). The most distant arc
(region 11) has an offset velocity relative to the arc second-scale
\HeII$\lambda$1640 emission from the AGN of $\sim$700\,\kms. We assumed
that the velocity of the \HeII\ line represents the systemic velocity of
the system. This assumption is supported by the small relative offset
velocity (\la 50\,\kms) between circum-nuclear dense gas (from \CO)
and \HeII\ in another radio galaxy \citep{gullberg16}.  Closer to the
nucleus, the filaments of \Lya\ emission narrow on the sky, the redshifted
offset velocity decreases to $\sim$250\,\kms, line full-width at half
maximum remain very high, reaching up to 700\,\kms, $\sim$50-100 kpc
from the nucleus. All throughout the emission, the lines are broad,
$\sim$300-700\,\kms.

Projected on to region 11, we find a z=3.1245 galaxy, which we have dubbed
the ``Arrow" because of its arrow-like morphology in HST/ACS F814W imaging
(Fig.~\ref{fig:imagespectralmontage}).

\section{What mechanisms power the \Lya\ emission?}\label{sec:heating}

\noindent
\textit{Photoionization by the AGN or star formation:} While the
rest-frame UV emission line ratios indicate that the AGN is likely
ionizing the gas within several 10s kpc along the direction of the
radio jets (regions 3-6), it is not clear if the AGN could plausibly
provide sufficient ionizing photons to explain the surface brightness
of the most extended gas. We can estimate the intensity of the
radiation field necessary to power the \Lya\ emission observed at
distances larger than $\sim$100 kpc. The surface brightness ranges
from $\sim$1-4 $\times$10$^{-19}$\,\surfb\ at such large distances
(3-10$\times$10$^{-17}$\,\surfb\ after correcting for surface
brightness dimming). Assuming case B recombination, the recombination
rate per unit area necessary to sustain these surface brightness is,
R$_\mathrm{rec}$= 4$\pi$I$_\mathrm{Ly\alpha}$/$\gamma_\mathrm{Ly\alpha}$,
where I$_\mathrm{Ly\alpha}$ is the intensity of the \Lya\ emission in
photons\,cm$^{-2}$\,s$^{-1}$\,sr$^{-1}$ and $\gamma_\mathrm{Ly\alpha}$
is the number of \Lya\ photons produced per recombination
(0.61). Assuming that every ionizing photon has an energy of 13.6
eV\footnote{The photo-ionization cross section of Hydrogen peaks
at 13.6 eV and this estimate is the maximum number of photons
expected. It is therefore an upper limit to the number of photons.}
and unity or greater covering fraction, we find the energy flux
necessary to sustain the range of surface brightness we observe is
$\approx$0.7-2$\times$10$^{-3}$\,erg\,cm$^{-2}$\,s$^{-1}$\,sr$^{-1}$.

We use H$\beta$ emission line luminosity as an indicator of the
photoionization rate from the AGN and massive stars.  The luminosity
of H$\beta$ \citep{nesvadba08} implies an ionizing luminosity,
L$_\mathrm{ion}$=4.8$\times$10$^{45}$\,erg\,s$^{-1}$. The
ionizing intensity of the AGN at a radius of 100 kpc is
$\approx$10$^{-3}$\,erg\,s$^{-1}$\,cm$^{-2}$\,sr$^{-1}$. The AGN provides
sufficient photons to explain the extended \Lya\ nebula within $\sim$100
kpc, but generally falls short for gas at larger distances. This estimate
only applies along lines-of-sight free of intervening absorption and
scattering, which may be true within the ionization cone, assumed to
lie along the radio axes with an opening angle $\sim$45\degr \citep[blue
regions in Fig.~\ref{fig:imagespectralmontage};][]{drouart12}.  The gas
along the radio jet, even the most extended gas along this direction
(regions 1 and 2) is thus plausibly to have its surface brightness
regulated by the AGN.  Away from the ionization cone (red regions 6--11
in Fig.~\ref{fig:imagespectralmontage}), it is unlikely that the AGN
provides sufficient photons; young stars embedded in these regions would
be an alternative. However, the remarkable uniformity of regions 6--10
suggests there is no significant local ionization by stars as observed
in, e.g., MRC\,0943-242 \citep{gullberg16} and also argues against a
central source such as an AGN ionizing the gas.

\noindent
\textit{Ionization by the meta-galactic flux:} At z$\sim$3,
the intensity of ionizing photons due to the meta-galactic
flux is $\approx$2$\times$10$^{5}$\,photons\,cm$^{-2}$\,s$^{-1}$
\citep{haardt96}. Assuming ionization equilibrium and the clouds are
optically thick, implies that the meta-galactic ionization rate is $\sim$2
orders-of-magnitude less than that necessary to maintain the ionization
of the extended gas.  So while the meta-galactic flux contributes to
the ionization of the diffuse gas, it does not maintain it.

\noindent
\textit{Resonance Scattering:} If the extended gas has sufficient
column of neutral HI to be optically thick at the wavelength of \Lya\,
then resonance scattering of the \Lya\ and UV continuum from the AGN
can contribute to its surface brightness.  The line profiles of the
extended \Lya\ emission do not mirror those of the nuclear gas -- they
are both narrower and have significant velocity offsets relative to
the nuclear emission (Table~\ref{tab:specnumbers}).  The approximately
constant surface brightness of the extended emission argues against
a central source exciting the emission. In addition, spectropolarimetry of 
\src\ \citep{reuland03} finds unpolarized ($<$4\%) \Lya\ emission, arguing 
against any significant contribution of scattered light to the circum-nuclear emission.

\noindent
\textit{Shock heating in the outer halo:} Another possibility for
exciting the \Lya\ emission in the halo of \src\ is an accretion shock
from inflowing gas at the halo boundary and/or shocks from cloud-cloud
collisions in a multiphase stream \citep{cornuault16}. The morphology
of the most extended emission to the west-southwest (region 11) is
shell-like and certainly suggestive of compression (a ``splash''). We
observe a velocity shear between region 11 and that of the inner halo
accross regions 6--10 of $\sim$400-500\,\kms. In addition, over these
regions, 6--11, the gas appears to be highly disturbed having FWHM of
300-700\,\kms.  These violent motions and shears make shock heating a
plausible mechanism for exciting the gas.

To test this possibility, we compare our measured surface brightness
with those predicted for high velocity shocks.  The models of
\citet{allen08} imply that even in relatively low density, low
metallicity, gas ($>$~few$\times$10$^{-3}$\,cm$^{-3}$), the surface
brightness produced by shocks is sufficient to explain our observations
for the relative velocities we observe in the
data, 300-700\,\kms. Simulations indicate filament densities of-order
10$^{-3}$ to 10$^{-1}$\,cm$^{-3}$ \citep{rosdahl12}, in agreement
with what is required. The surface brightness is related to the
mass flow rate. These densities and velocities, imply a mass flow rate,
$\rho_\mathrm{pre-shock}$\,v$_\mathrm{shock}$\,A$_\mathrm{shock}$\,$\approx$250\,\Msun\,yr$^{-1}$,
for pre-shock density, $\rho_\mathrm{pre-shock}$\,=\,0.01\,cm$^{-2}$,
shock velocity, v$_\mathrm{shock}$\,=\,500\,\kms, and area,
A$_\mathrm{shock}$, and for simplicity, of a circle with a radius
of 25\,kpc.

While we have used the observed high relative velocities and large
line widths as a justification of exploring the possibility of shock
heating, caution is warranted. Both the relative velocities and
large line widths may be partially due to radiative transfer effects
\citep{cantalupo05}. The size of the effect depends on whether or not
the medium is multiphasic and what causes the ionization.  However, none
of the \Lya\ lines in the most extended regions (1, 2, and 6--11) appear
split as might be expected if radiative transfer effects are important.

\section{\Lya\ structures as accretion streams}\label{sec:ifacc}

Comparing the relative velocities of galaxies surrounding \src\ with
those from cosmological simulations, \citet{kuiper12} suggest that it is
a result of the on-going merger of two massive, $\sim$10$^{14}$\,\Msun
, proto-clusters. Finding that the surface brightness of the gas in
the outer halo is consistent with being shocked heated, its arc- or
shell-like and filamentary morphology, large line widths, redshifted
velocities, and relatively smooth change in velocity as a function
of distance from the AGN suggests this may be gas accreting from the
cosmic web \citep{rosdahl12, goerdt15a}. If we make this assumption,
then what do our results tell us about gas accretion into massive halos
and do they agree with those from cosmological simulations?

For this comparison, we focus exclusively on the filamentary structure to
the west-southwest of the AGN (regions 6--11). We limit ourselves to these
regions because none of the low surface brightness emission shows obvious
evidence for being ionized by the AGN, they do not lie along the direction
of the radio jets \citep[which trace the direction of the ionization cone
and the ionization cones have opening angles $\sim$45\degr, suggesting
these regions lie outside;][]{drouart12}, and they extend rather
continuously in emission from the outer halo into the circum-nuclear
region. As already noted in the introduction, these regions have surface
brightnesses consistent with estimates from simulations \citep{rosdahl12}.
The ``Arrow'' lies in this structure which is seen in simulated
filaments \citep[][Fig.~\ref{fig:imagespectralmontage}]{gheller16}.
These observations form the basis for us assuming these regions are part
of an accretion stream.

The surface brightness of the distant arc can be explained via shock
heating. Assuming such shock occurs at the virial radius, simple
scaling relations imply radii of $\sim$150 and 330 kpc for 10$^{13}$ and
10$^{14}$\,\Msun\ halos respectively. Similarly, the virial velocities for
halos with these masses are 500 and 1100\,\kms . We do not necessarily
expect the velocities of the streams to be equal to the virial velocity
but can be as low as half \citep{goerdt15b}. Both the distance of the
outermost shell from the nucleus and the velocities we observe are
consistent with virial expectations. We observe projected distances
and velocities (and the projection of the velocity could change with
position), not true positions and velocities, so this comparison is, by
necessity, only order-of-magnitude. Simulations do show broad lines in
Ly$\alpha$, of the same order (100s\,\kms) as we observe \citep{goerdt10}.

Gas accretion through a stream, while able to make up most of the total 
accretion rate, can also vary substantially within a filament. \citet{goerdt15a} 
found accretion rates of $\sim$50-5000\,\Msun\,yr$^{-1}$\,sr$^{-1}$ through filaments for a
halo mass of $\approx$10$^{12}$\,\Msun. Scaling up these rates per sr
by the virial mass as is appropriate for total mass accretion rates
\citep[$\dot{\textrm{M}}\,\propto\,\textrm{M}^{1.25}_\textrm{vir}$,][]{goerdt15a}, 
we estimate rates 15 to 300$\times$ higher for 10 to 100$\times$ higher halo masses. 
A rough estimate suggests the shock subtends
$\sim$1\% of the spherical surface area at 250\,kpc and using the mass
flow rate (\S~\ref{sec:heating}), implies an accretion rate per
unit solid angle $\sim$2$\times$10$^{4}$\,\Msun\,yr$^{-1}$\,sr$^{-1}$
-- a value consistent with simulations.

Many properties of the \Lya\ emission we observed in the circum-galactic
medium of \src\ are broadly consistent with the results of simulations
of gas accretion streams. However, we view aspects of this agreement as
fortuitous. Simulations generally lack the spatial and temporal resolution
necessary to capture thermal and dynamical instabilities, resolve high
Mach shock fronts, fragmentation of the post-shock gas, and turbulence
that naturally occur in astrophysical gas flows \citep{kritsuk02,
cornuault16}. It is not clear if simulations show conspicuously large
shock fronts and strong post-shock cooling at approximately the halo
boundary as we have suggested.  Such fronts, if real, are important for
establishing many properties of the gas on its inward journey deeper into
the potential well (fragmenting the gas and inducing turbulence).  Until
simulations reach the necessary resolutions and input physics to resolve
shock fronts and fragmentation, it is fair to say that our theoretical
understanding of these flows is limited. Observational constraints, such
as those provided here, are crucial for obtaining a deeper understanding
of how galaxies get their gas through gas accretion flows.

\begin{acknowledgements}

MDL thanks the ESO visitor program for support.

\end{acknowledgements}

\bibliographystyle{aa}
\bibliography{HzRG_bib}

\end{document}

%% file: Results_fit.tex
\begin{table}
\centering
	{\small
\caption{Characteristics of spectra of particular regions}
\label{tab:specnumbers}
\begin{tabular}{c c c c c}
\hline \hline
\textit{Id} & SB $\times$ 10$^{-19}$  & $\Delta$ V & FWHM & Area \\ 
	& {\small erg s$^{-1}$ cm$^{-2}$ arcsec$^{-2}$} & {\small km s$^{-1}$} & {\small km s$^{-1}$} & {\small arcsec$^2$} \\
\hline
 1    &   1.3$\pm$0.2  &    142$\pm$13   &  250$\pm$32  & 14.3 \\
 2    &   1.6$\pm$0.1  &    220$\pm$9   &  430$\pm$21  & 46.5 \\
 3    &   1.4$\pm$0.1  &    211$\pm$12   &  550$\pm$29  & 40.5 \\
 4    &   5.5$\pm$0.2  &    224$\pm$5   &  410$\pm$13  & 9.3 \\
 5    &   2.8$\pm$0.1  &    225$\pm$11   &  550$\pm$26  & 11.4 \\
 6    &   2.1$\pm$0.1  &    230$\pm$6   &  395$\pm$16  & 43.4 \\
 7    &   0.8$\pm$0.1  &    227$\pm$25   &  525$\pm$62  & 35.0 \\
 8    &   0.9$\pm$0.1  &    226$\pm$28   &  670$\pm$70  & 17.2  \\
 9    &   2.6$\pm$0.2  &    250$\pm$14   &  510$\pm$32  & 8.5  \\
10    &   0.8$\pm$0.1  &    290$\pm$13   &  510$\pm$32  & 74.0  \\
11    &   1.9$\pm$0.1  &    610$\pm$11   &  720$\pm$27  & 36.6  \\
\hline
\end{tabular}
\tablefoot{The identification numbers correspond to the regions in
	Fig.~\ref{fig:imagespectralmontage}. The measured $\Delta$V of regions 2-9 are statistically the same.}}
\end{table}